\documentclass[[aps,twocolumn,showpacs,amsmath,amssymb]{revtex4-1}

\usepackage{epsfig}
\usepackage{graphicx}
\usepackage{booktabs}
\usepackage{multirow}
\usepackage{tabularx}
\usepackage{times}
\usepackage{color}
\usepackage{subfigure}
\usepackage{todonotes}
\usepackage{ulem}
\usepackage[english]{babel}

\definecolor{Red}{rgb}{1,0,0}

\definecolor{Blu}{rgb}{0,0,01}

\definecolor{Green}{rgb}{0,1,0}

\newcommand{\be}{\begin{equation}}
\newcommand{\ee}{\end{equation}}

\begin{document}

\title{Magnetization reorientation due to the superconducting transition in heavy-metal heterostructures}

\author{Lina G. Johnsen}

\affiliation{Center for Quantum Spintronics, Department of Physics, Norwegian
University of Science and Technology, NO-7491 Trondheim, Norway}

\author{Niladri Banerjee}

\affiliation{Department of Physics, Loughborough University, Loughborough,
LE11 3TU, United Kingdom}

\author{Jacob Linder}

\affiliation{Center for Quantum Spintronics, Department of Physics, Norwegian
University of Science and Technology, NO-7491 Trondheim, Norway
\bigskip}

\begin{abstract}
Recent theoretical and experimental work has demonstrated how the superconducting critical temperature $(T_c )$ can be modified by rotating the magnetization of a single homogeneous ferromagnet proximity-coupled to the superconducting layer. This occurs when the superconductor and ferromagnet are separated by a thin heavy normal metal that provides an enhanced interfacial Rashba spin-orbit interaction. In the present work, we consider the reciprocal effect: magnetization reorientation driven by the superconducting phase transition. We solve the tight-binding Bogoliubov-de Gennes equations on a lattice self-consistently and compute the free energy of the system. We find that the relative angle between the spin-orbit field and the magnetization gives rise to a contribution in the free energy even in the normal state, $T>T_c$, due to band-structure effects. For temperatures below $T_c$, superconductivity gives rise to a competing contribution. We demonstrate that by lowering the temperature, in addition to reorientation of the favored magnetization direction from in-plane to out-of-plane, a $\pi/4$ in-plane rotation for thicker ferromagnetic layers is possible. Furthermore, computation of $T_c$ of the structure in the ballistic limit shows a dependence on  the in-plane orientation of the magnetization, in contrast to our previous result on the diffusive limit. This finding is relevant with respect to thin-film heterostructures since these are likely to be in the ballistic regime of transport rather than in the diffusive regime. Finally, we discuss the experimental feasibility of observing the magnetic anisotropy induced by the superconducting transition when other magnetic anisotropies, such as the shape anisotropy for a ferromagnetic film, are taken into account. Our work suggests that the superconducting condensation energy in principle can trigger a reorientation of the magnetization of a thin-film ferromagnet upon lowering the temperature below $T_c$, in particular for ferromagnets with weak magnetic anisotropies.
\end{abstract}

\date{\today}

\maketitle

\section{Introduction}

Recent research within the field of superconducting spintronics has focused on combining superconducting and magnetic materials into hybrid structures to study novel phases arising from proximity effects not found in individual materials \cite{Linder2015Superconducting}. In conventional superconductors (S), Cooper pairs exists as spin-singlet pairs. The two electrons in a pair have opposite spin and are destroyed when they enter a ferromagnet (F) as they quickly lose their coherence due to the magnetic exchange field. At the interface between a superconductor and a ferromagnet, spin-singlet pairs are transformed into spin-zero triplet Cooper pairs that have a short penetration depth into the ferromagnetic region. However, two misoriented ferromagnets breaking spin-rotational symmetry can transform opposite-spin triplets into equal-spin triplets \cite{Eschrig2015Spin}. Due to their equally directed spins along the magnetization direction, these Cooper pairs maintain coherence longer and are instead able to survive for a longer distance inside the ferromagnet. The density of equal-spin triplets in the system depends on the relative orientation of the ferromagnets \cite{Eschrig2015Spin,Eschrig2011Spin}. This has been demonstrated experimentally (see for instance Refs. \cite{Gu2002Magnetization, Moraru2005Magnetization, Leksin2012Evidence, banerjee_natcom_14, wang_prb_14}) by showing a variation of $T_c$ in a F1/S/F2 or F1/F2/S system by changing the relative magnetization of the F1 and F2 layers.
This variation is attributed to the generation of triplet Cooper pairs with increasing misalignment of the magnetizations of F1 and F2 layer moments. Recent research \cite{Jacobsen2015Critical, Ouassou2016Electric,simensen_prb_18, Banerjee2017Controlling} has reported a similar modulation of the critical temperature by changing the orientation of a single homogeneous ferromagnet coupled to a superconductor through a thin heavy normal metal (HM) film with strong Rashba spin-orbit coupling. Measurements \cite{Banerjee2017Controlling} performed on a Nb/Pt/Co/Pt system showed a suppression of the critical temperature for an in-plane (IP) magnetization that was attributed to a reduced superconducting gap due to triplet generation. A reduced gap also implies an increase in the free energy since part of the superconducting condensation energy is lost. We may therefore suspect the superconducting contribution to the free energy to favor an out-of-plane (OOP) magnetization direction.

Motivated by this, here we explore the striking possibility of reorienting the magnetization of the ferromagnetic layer in an S/HM/F system by changing the temperature. We discover that upon lowering the temperature below $T_c$, the dependence of the free energy on the magnetization direction changes due to the superconducting phase transition. In turn, this leads to a change in the ground state magnetization direction, or effectively the magnetization angle that minimizes the free energy. For sufficiently thin ferromagnetic layers, we get a change from IP to OOP magnetization. We also find that there is an IP variation in the free energy and show that it is in principle possible to get an IP $\pi/4$ rotation of the magnetization when lowering the temperature below $T_c$. This opens the possibility for temperature-induced switching of the magnetization both between the IP and OOP orientation and switching within a plane parallel to the interface.

\section{Theory}

To describe our S/HM/F system, we use the tight-binding Bogoliubov-de Gennes (BdG) framework and use conventions similar to those in Refs. \cite{Terrade2015Proximity,Linder2017Intrinsic}. The lattice BdG framework is well suited for describing heterostructures, fully accounts for the crystal symmetry of the electronic environment, and can describe atomically thin layers of a material. The Hamiltonian we use is
\begin{equation}
    \begin{split}
        \label{Hamiltonian}
        &H = -t\sum_{\left<\boldsymbol{i},\boldsymbol{j}\right>,\sigma}c_{\boldsymbol{i},\sigma}^\dagger c_{\boldsymbol{j},\sigma}
        -\sum_{\boldsymbol{i},\sigma} \mu_{\boldsymbol{i}} c_{\boldsymbol{i},\sigma}^\dagger c_{\boldsymbol{i},\sigma}
        -\sum_{\boldsymbol{i}} U_{\boldsymbol{i}}n_{\boldsymbol{i},\uparrow}n_{\boldsymbol{i},\downarrow} \\
        &\hspace{8.5mm}-\frac{i}{2}\sum_{\left< \boldsymbol{i},\boldsymbol{j}\right> ,\alpha,\beta} \lambda_{\boldsymbol{i}} c_{\boldsymbol{i},\alpha}^{\dagger} \hat{n} \cdot (\boldsymbol{\sigma}\times\boldsymbol{d}_{\boldsymbol{i},\boldsymbol{j}})_{\alpha,\beta} c_{\boldsymbol{j},\beta}\\
        &\hspace{8.5mm}+\sum_{\boldsymbol{i},\alpha,\beta}c_{\boldsymbol{i},\alpha}^{\dagger}(\boldsymbol{h}_{\boldsymbol{i}}\cdot\boldsymbol{\sigma})_{\alpha,\beta} c_{\boldsymbol{i},\beta}
    \end{split}
\end{equation}
Above, $t$ is the hopping integral, $\mu_{\boldsymbol{i}}$ is the chemical potential at lattice site $\boldsymbol{i}$, $U>0$ is the attractive on-site interaction that gives rise to superconductivity, $\lambda_{\boldsymbol{i}}$ is the Rashba spin-orbit coupling magnitude at site $\boldsymbol{i}$, $\hat{n}$ is a unit vector normal to the interface, $\boldsymbol{\sigma}$ is the vector of Pauli matrices, $\boldsymbol{d}_{\boldsymbol{i},\boldsymbol{j}}$ is the vector from site $\boldsymbol{i}$ to site $\boldsymbol{j}$, and $\boldsymbol{h_i}$ is the local magnetic exchange field.  $c_{\boldsymbol{i},\sigma}^\dagger$ and $c_{\boldsymbol{i},\sigma}$ are the second-quantization electron creation and annihilation operators at site $\boldsymbol{i}$ with spin $\sigma$, and $n_{\boldsymbol{i},\sigma}\equiv c_{\boldsymbol{i},\sigma}^\dagger c_{\boldsymbol{i},\sigma}$. The superconducting term in the Hamiltonian is treated by a mean-field approach, where we insert $c_{\boldsymbol{i},\uparrow} c_{\boldsymbol{i},\downarrow} = \left< c_{\boldsymbol{i},\uparrow} c_{\boldsymbol{i},\downarrow} \right> +\delta$ and $c_{\boldsymbol{i},\uparrow}^\dagger c_{\boldsymbol{i},\downarrow}^\dagger = \big< c_{\boldsymbol{i},\uparrow}^\dagger c_{\boldsymbol{i},\downarrow}^\dagger \big> +\delta^\dagger$ into Eq. \ref{Hamiltonian} and neglect terms of second order in the fluctuations $\delta$ and $\delta^\dagger$. $\Delta_{\boldsymbol{i}} \equiv U_{\boldsymbol{i}}\left<c_{\boldsymbol{i},\uparrow}c_{\boldsymbol{i},\downarrow}\right>$ is the superconducting order parameter, which we solve for self-consistently.
\begin{figure}[h]
    \centering
    \includegraphics[width=0.8\columnwidth]{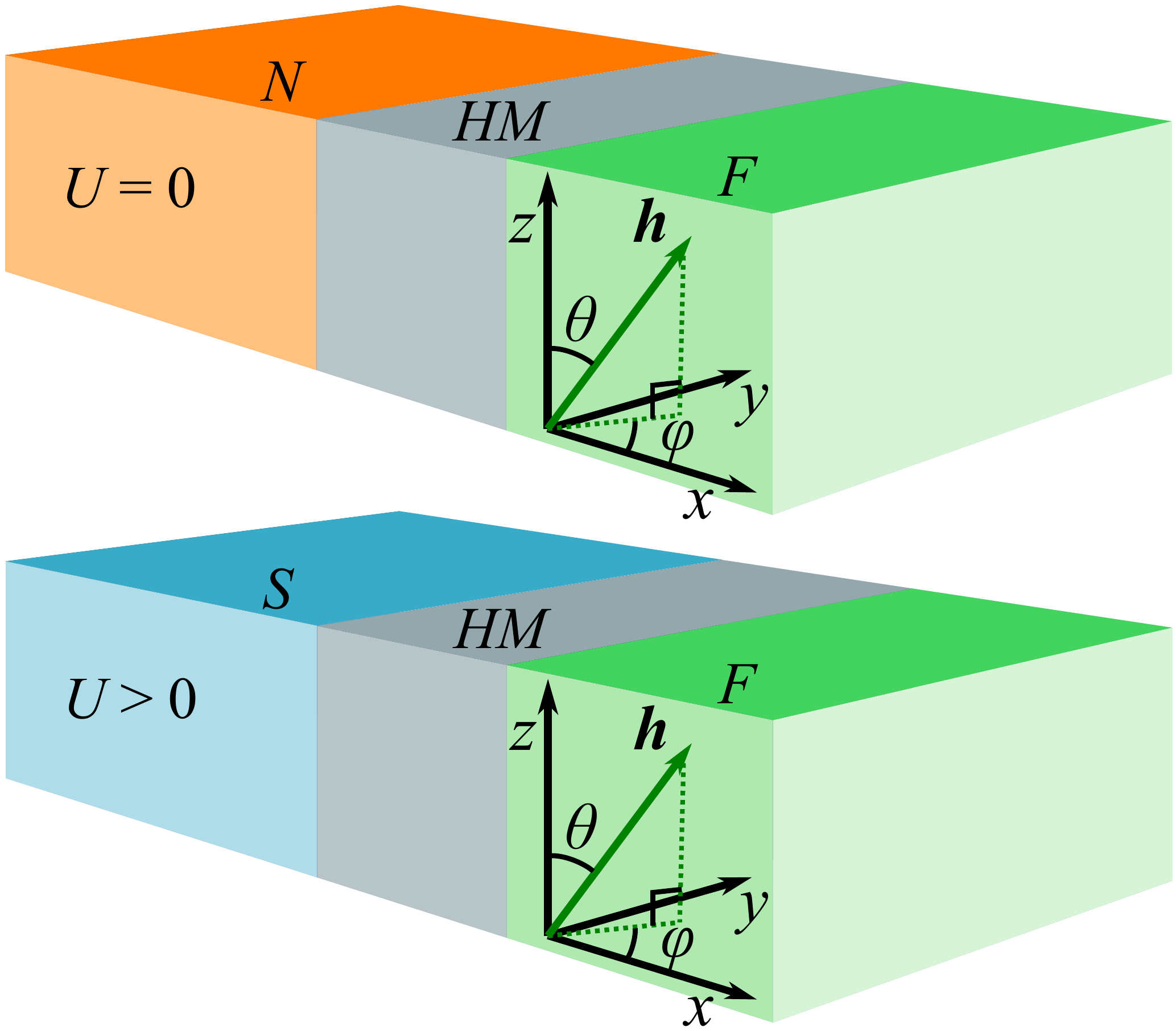}
    \caption{Suggested experimental setup for demonstrating a magnetization reorientation due to a change in temperature. We have a stack of a normal-metal layer ($T>T_c $, $U=0$) or a superconducting layer ($T<T_c$, $U>0$), a heavy-metal layer, and a ferromagnetic layer. We model our system as a 3D cubic lattice with interface normal along the $x$ direction. The exchange field $\boldsymbol{h}$ is described by the polar and azimuthal angles with respect to the $z$ axis, $(\theta,\phi)$. Note that the above model is not to scale.}
    \label{fig:model}
\end{figure}
We consider a 3D cubic lattice of size $N_x \times N_y \times N_z$, as shown in Fig. \ref{fig:model}. The lattice consists of three layers: a superconducting layer, a thin heavy-metal layer with Rashba spin-orbit coupling, and a thin ferromagnetic layer. For enabling experimental observation of the effects considered in this paper, the system should have as good an interface quality as possible to maximize the proximity effect, and heavy metal interlayers should be used to boost the spin-orbit coupling strength. For concrete material choices, we suggest a Nb superconductor with Pt interlayers, which should give a strong proximity effect and strong spin-orbit coupling (see for instance Ref. \cite{satchell_prb_18}). In addition, the ferromagnet should be soft and have as weak an anisotropy as possible. We suggest using a $7\%$ Mo-doped permalloy, which has a very low switching energy \cite{gingrich_thesis_14}. We describe the trilayer system shown in Fig. \ref{fig:model} using the Hamiltonian in Eq. \ref{Hamiltonian}, where the terms are only nonzero in their respective regions. The interface normals are parallel to the $x$ axis $(\hat{n}=\hat{x})$. We assume periodic boundary conditions in the $y$ and $z$ directions, so that all quantities depend on the $x$ component of the site index only. In our presentation of the results, we scale all energies to the hopping element, $t$, and all lengths to the lattice constant, $a$. For simplicity, we also set the reduced Planck constant $\hbar$ and the Boltzmann constant $k_B$ equal to 1. Therefore, all temperatures are scaled by $t/k_B$ in the presentation of the results. The magnetic exchange field of the ferromagnet is expressed by $\textbf{h}=h(\cos(\phi)\sin(\theta),\sin(\phi)\sin(\theta),\cos(\theta))$, where $\theta$ is the polar angle with respect to the $z$-axis and $\phi$ is the azimuthal angle. Because of our assumption of periodic boundary conditions along $\hat{y}$ and $\hat{z}$, the Fourier transform
\begin{equation}
    \label{FT}
    c_{\boldsymbol{i},\sigma}=\frac{1}{\sqrt{N_y N_z}}\sum_{k_y,k_z} c_{i_x,k_y,k_z,\sigma} e^{i(k_y i_y +k_z i_z)}.
\end{equation}
can be used to diagonalize the Hamiltonian.
The sum is over the allowed $k_y$ and $k_z$ inside the first Brillouin zone. Also note that
\begin{equation}
\begin{split}
    \label{rel}
    &\frac{1}{\sqrt{N_y}}\sum_{i_y} e^{i(k_y-k_y ')i_y}=\delta_{k_y , k_y '},\\
    &\frac{1}{\sqrt{N_z}}\sum_{i_z} e^{i(k_z-k_z ')i_z}=\delta_{k_z , k_z '}.
\end{split}
\end{equation}
We choose a basis 
\begin{equation}
\begin{split}
    &B_{i_x , k_y , k_z}^{\dagger}\\
    &=[c_{i_x , k_y , k_z ,\uparrow}^{\dagger} \hspace{3mm} c_{i_x , k_y , k_z ,\downarrow}^{\dagger} \hspace{3mm} c_{i_x , -k_y , -k_z ,\uparrow} \hspace{3mm} c_{i_x , -k_y , -k_z ,\downarrow}]
\end{split}
\end{equation} 
and rewrite the Hamiltonian as
\begin{equation}
\label{H2}
    H=H_0+\frac{1}{2}\sum_{i_x , j_x , k_y , k_z} B_{i_x , k_y , k_z}^\dagger H_{i_x , j_x , k_y , k_z}B_{i_x , k_y , k_z}.
\end{equation}
By using Eqs. \ref{FT} and \ref{rel} to rewrite the Hamiltonian in Eq. \ref{Hamiltonian}, we can show that
\begin{equation}
\label{Hamiltonian2}
    \begin{split}
        &H_{i_x , j_x , k_y , k_z}= \epsilon_{i_x ,j_x ,k_y ,k_z} \hat{\tau}_3 \hat{\sigma}_0\\
        &\hspace{17.5mm}+\delta_{i_x ,j_x}\Big[h_{i_x}^x \hat{\tau}_3 \hat{\sigma}_x +h_{i_x}^y \hat{\tau}_0 \hat{\sigma}_y +h_{i_x}^z \hat{\tau}_3 \hat{\sigma}_z\\
        &\hspace{17.5mm} -\lambda_{i_x}\sin(k_y )\hat{\tau}_0 \hat{\sigma}_z +\lambda_{i_x}\sin(k_z )\hat{\tau}_3 \hat{\sigma}_y\\
        &\hspace{17.5mm}+\Delta_{i_x}i\hat{\tau}^+ \hat{\sigma}_y  -\Delta_{i_x}^\star i\hat{\tau}^- \hat{\sigma}_y  \Big],\\
    \end{split}
\end{equation}
where
\begin{equation}
\begin{split}
    &\epsilon_{i_x , j_x , k_y , k_z} \equiv -2t\left(\cos(k_y)+\cos(k_z)\right)\delta_{i_x , j_x}\\
    &\hspace{19mm}-t(\delta_{i_x , j_x +1}+\delta_{i_x , j_x -1})-\mu_{i_x}\delta_{i_x , j_x}
\end{split}
\end{equation}
and $\hat{\tau}^\pm =(\hat{\tau}_1 \pm i\hat{\tau}_2 )/2$. Above, $\hat{\tau}_i \hat{\sigma}_j \equiv \hat{\tau}_i \otimes\hat{\sigma}_j$ is the Kronecker product of the Pauli matrices spanning Nambu and spin space.  The constant term is
\begin{equation}
\begin{split}
    &H_0=N_y N_z \sum_{i_x}\frac{|\Delta_{i_x}|^2}{U_{i_x}} \\
    &\hspace{8mm}- \sum_{i_x , k_y , k_z}\left[2t\left(\cos(k_y)+\cos(k_z)\right)+\mu_{i_x}\right].
\end{split}
\label{H0}
\end{equation}
By defining another basis,
\begin{equation}
    W_{k_y , k_z}^\dagger = [B_{1,k_y ,k_z}^\dagger,...,B_{i_x ,k_y ,k_z}^\dagger ,...,B_{N_x ,k_y ,k_z}^\dagger ],
\end{equation}
Eq. \ref{H2} can be rewritten as 
\begin{equation}
    H=H_0 + \frac{1}{2}\sum_{k_y , k_z}W_{k_y ,k_z}^\dagger H_{k_y , k_z} W_{k_y , k_z},
\end{equation}
where
\begin{equation}
\label{Hkykz}
    H_{k_y ,k_z}=
    \begin{bmatrix}
        H_{1,1,k_y ,k_z} & \cdots & H_{1,N_x ,k_y ,k_z}\\
        \vdots & \ddots & \vdots \\
        H_{N_x ,1,k_y ,k_z} & \cdots & H_{N_x ,N_x ,k_y ,k_z}\\
    \end{bmatrix}.
\end{equation}
$H_{k_y ,k_z}$ is Hermitian and can be diagonalized numerically with eigenvalues $E_{n,k_y ,k_z}$ and eigenvectors $\Phi_{n,k_y ,k_z}$ given by 
\begin{equation}
\begin{split}
    &\Phi_{n,k_y ,k_z}^{\dagger}=[ \phi_{1,n,k_y ,k_z}^{\dagger} \hspace{3mm} \cdots \hspace{3mm}\phi_{N_x ,n,k_y ,k_z}^{\dagger}],\\
    &\phi_{i_x ,n,k_y ,k_z}^{\dagger}=[u_{i_x ,n,k_y ,k_z}^{\star}\hspace{1mm} v_{i_x ,n,k_y ,k_z}^{\star}\hspace{1mm} w_{i_x ,n,k_y ,k_z}^{\star}\hspace{1mm} x_{i_x ,n,k_y ,k_z}^{\star}].
\end{split}
\end{equation} 
The diagonalization is done numerically and gives a Hamiltonian of the form
\begin{equation}
    H=H_0+\frac{1}{2}\sum_{n, k_y , k_z}E_{n, k_y , k_z}\gamma_{n, k_y , k_z}^\dagger \gamma_{n, k_y , k_z},
\end{equation}
where the new quasiparticle operators are related to the old operators by
\begin{equation}
\label{c}
\begin{split}
    & c_{i_x ,k_y ,k_z ,\uparrow}=\sum_n u_{i_x ,n, k_y ,k_z}\gamma_{n,k_y ,k_z},\\
    &c_{i_x ,k_y ,k_z ,\downarrow}=\sum_n v_{i_x ,n, k_y ,k_z}\gamma_{n,k_y ,k_z},\\
    &c_{i_x ,-k_y ,-k_z ,\uparrow}=\sum_n w_{i_x ,n, k_y ,k_z}\gamma_{n,k_y ,k_z},\\
    &c_{i_x ,-k_y ,-k_z ,\downarrow}=\sum_n x_{i_x ,n, k_y ,k_z}\gamma_{n,k_y ,k_z}.
\end{split}
\end{equation}
To find the eigenvectors and eigenvalues the initial guess of the order parameter must be improved by iterative treatment. The expression for the gap can be rewritten by inserting the operators given in Eq. \ref{c} and by using that $\langle \gamma_{n,k_y ,k_z}^\dagger \gamma_{m,k_y ,k_z}\rangle =f\big(E_{n,k_y ,k_z}/2\big)\delta_{n,m}$. We get
\begin{equation}
    \label{deltaIt}
    \begin{split}
        &\Delta_{i_x}=\\
        &-\frac{U_{i_x}}{N_y N_z}\sum_{n,k_y , k_z}v_{i_x ,n,k_y ,k_z}w_{i_x ,n,k_y ,k_z}^\star \left[1-f\left(E_{n,k_y ,k_z}/2\right)\right].
    \end{split}
\end{equation}
Here, $f\big(E_{n,k_y ,k_z}/2\big)$ is the Fermi-Dirac distribution. 

Having found $E_{n,k_y ,k_z}$ and $\{u,v,w,x\}$, we can compute the physical quantities of interest. The free energy is given by
\begin{equation}
    \label{F}
    F=H_0 -\frac{1}{\beta}\sum_{n,k_y ,k_z}\ln(1+e^{-\beta E_{n, k_y ,k_z}/2}),
\end{equation}
where $\beta=(k_B T)^{-1}$. Note that if $T\to 0$,
\begin{equation}
    \label{FT->0}
    F=H_0 +\frac{1}{2}{\sum_{n,k_y ,k_z}}^{'} E_{n,k_y ,k_z},
\end{equation}
where $\sum^{'}$ means that the sum is taken over negative eigenenergies only. The ground state of the system minimizes the free energy. $F$ is therefore used to find the preferred orientation of the ferromagnet. Additional magnetic anisotropy terms may be added to the free energy to take the thickness of the thin ferromagnetic film into account more properly. We model these terms in a simple way and write the additional contribution to the free energy as \cite{Johnson1996Magnetic}
\begin{equation}
    \label{anis}
    F_a = -K_{\text{eff}}\cos^2(\theta_p).
\end{equation}
where $\theta_p$ is the polar angle relative to the interface normal.  $K_{\text{eff}}$ is the effective anisotropy constant. We assume a thin ferromagnetic film with one interface to another material and one free surface, and approximate $K_{\text{eff}}$ by \cite{Johnson1996Magnetic}
\begin{equation}
\label{Keff}
K_{\text{eff}} = K_v + \frac{K_s + K_i}{t_F}.
\end{equation}
Above, $K_v<0$ is the bulk anisotropy of the ferromagnet, $K_s$ is the surface anisotropy and $K_{i}>0$ is the anisotropy of the interface between the ferromagnet and the other material. $K_{\text{eff}}$ may be positive or negative depending on the thickness of the ferromagnetic layer, $t_F$. If $K_{\text{eff}}<0$, the magnetic anisotropy contribution $F_a$ to the free energy favors IP magnetization and shape anisotropy dominates. For $K_{\text{eff}}>0$, OOP magnetization is favored and perpendicular anisotropy dominates. To model a non-cubic ferromagnet, we use the average lattice constant, $a=(a_x +a_y +a_z )/3$. By doing this we obtain a rather rough estimate of $F_a$, but since we are comparing $F_a$ to the superconducting contribution to the free energy, the order of magnitude of the change in $F_a$ is more interesting than the details.

The physical mechanism leading to a variation in the superconducting condensation energy when the magnetization direction changes is the conversion of singlet Cooper pairs to triplet ones. To reveal the types of triplet Cooper pairs in our system, we compute the triplet anomalous Green's function amplitudes. The on-site odd-frequency $s$-wave anomalous triplet amplitudes are defined as
\begin{equation}
\begin{split}
\label{S}
    &S_{0,\boldsymbol{i}}(\tau)=\left<c_{\boldsymbol{i},\uparrow}(\tau)c_{\boldsymbol{i},\downarrow}(0)\right>+\left<c_{\boldsymbol{i},\downarrow}(\tau)c_{\boldsymbol{i},\uparrow}(0)\right>,   \\ &S_{\sigma,\boldsymbol{i}}(\tau)=\left<c_{\boldsymbol{i},\sigma}(\tau)c_{\boldsymbol{i},\sigma}(0)\right>,
\end{split}
\end{equation}
where we have defined the time-dependent electron annihilation operator $c_{\boldsymbol{i},\sigma}(\tau)\equiv e^{iH\tau}c_{\boldsymbol{i},\sigma}e^{-iH\tau}$. By differentiating $c_{\boldsymbol{i},\sigma}(\tau)$ with respect to $\tau$ we obtain the Heisenberg equation,
\begin{equation}
\label{Heisenberg}
    \frac{dc_{\boldsymbol{i},\sigma}(\tau)}{d\tau}=i[H,c_{\boldsymbol{i},\sigma}(\tau)],
\end{equation}  
from which we can obtain expressions for $c_{\boldsymbol{i},\uparrow}(\tau)$ and $c_{\boldsymbol{i},\downarrow}(\tau)$ by inserting Eq. \ref{c}. Here, $\tau$ is the relative time coordinate between the electron operators. $\tau$ is scaled by $\hbar/t$. The even-frequency p-wave anomalous triplet amplitudes are defined
\begin{equation}
\begin{split}
\label{P} 
    &P_{0,\boldsymbol{i}}^{n} =\sum_\pm \pm(\left<c_{\boldsymbol{i},\uparrow}c_{\boldsymbol{i}\pm\hat{n},\downarrow}\right>+\left<c_{\boldsymbol{i},\downarrow}c_{\boldsymbol{i}\pm\hat{n},\uparrow}\right>),\\
    &P_{\sigma,\boldsymbol{i}}^{n} =\sum_\pm \pm\left<c_{\boldsymbol{i},\sigma}c_{\boldsymbol{i}\pm\hat{n},\sigma}\right>,
\end{split}
\end{equation}
    where $n=\{x,y,z\}$. The spins in these triplet amplitudes are defined with respect to the $z$ axis. If we want to compute the triplet amplitudes for a specific direction of $\boldsymbol{h}$ such that $(\uparrow\uparrow)_{\boldsymbol{h}}$ and $(\downarrow\downarrow)_{\boldsymbol{h}}$ represent the long-range triplets, the triplet amplitudes must be transformed so that the spins are defined with respect to the vector $\boldsymbol{h}$. This is done by inserting  $(c_{\boldsymbol{i},\uparrow})_{\theta, \phi} = \cos(\theta/2)e^{-i\phi/2}(c_{\boldsymbol{i},\uparrow})_z +\sin(\theta/2)e^{i\phi/2}(c_{\boldsymbol{i},\downarrow})_z$ and $(c_{\boldsymbol{i},\downarrow})_{\theta, \phi} = -\sin(\theta/2)e^{-i\phi/2}(c_{\boldsymbol{i},\uparrow})_z +\cos(\theta/2)e^{i\phi/2}(c_{\boldsymbol{i},\downarrow})_z$ \cite{Eschrig2015Spin} into Eqs. \ref{S} and \ref{P}. The even-frequency $s$-wave singlet amplitude is proportional to the gap and given by
\begin{equation}
    S_{s,\boldsymbol{i}}=\left<c_{\boldsymbol{i},\uparrow}c_{\boldsymbol{i},\downarrow}\right>-\left<c_{\boldsymbol{i},\downarrow}c_{\boldsymbol{i},\uparrow}\right>.
\end{equation}
The singlet amplitude is rotationally invariant with respect to the choice of quantization axis, and the quantity
\begin{equation}
    \Tilde{S}_s=\frac{1}{N_{x,S}}\sum_{i_x}|S_{s,i_x}|
\end{equation}
is a measure of the singlet amplitude of the system for a given magnetization direction. The sum is taken over the superconducting region only, as we are primarily interested in describing how the superconducting condensation energy depends on the magnetization direction.

We find $T_c$ numerically by a binomial search within temperatures below the bulk critical temperature of the superconductor. In each of the $n$ iterations, we determine whether $T_c$ is above or below the temperature in the middle of the current temperature interval. This is done by choosing an initial guess for $\Delta_{i_x}$ very close to zero and checking whether $\Delta_{i_x} (T)$ close to $N_{x,S} /2$ increases or decreases from the initial guess after recalculating $\Delta_{i_x}$ $m$ times by Eq. \ref{deltaIt}. The gap decreases in the normal state and increases in the superconducting state.

The superconducting coherence length $(\xi)$ of the superconducting layer is an important length scale in our system. The effects of the HM/F layer can be expected to be strongest when $\xi$ is the same length or slightly longer than the thickness of the superconductor. In the ballistic limit the superconducting coherence length is given by $\xi=\hbar v_F /\pi\Delta_0$ \cite{Bardeen1957Theory}. The normal-state Fermi velocity, $v_F$, is obtained by the dispersion relation  $v_F=\frac{1}{\hbar}\frac{dE_{\boldsymbol{k}}}{dk}\big|_{k=k_F}$ \cite{Bardeen1957Theory}. $E_{\boldsymbol{k}}=-2t[\cos(k_x )+\cos(k_y )+\cos(k_z )]-\mu_N$ is the normal state eigenenergies obtained from Eq. \ref{Hamiltonian} if we use periodic boundary conditions in all three directions. The Fermi momentum $k_F$ corresponds to the Fermi energy, which is the highest occupied energy level at $T=0$. $\Delta_0$ is the zero-temperature superconducting gap. In our lattice model we round $\xi$ down to the closest integer number of lattice points.

In our calculations, we have used a 3D cubic lattice model with periodic boundary conditions in both the $y$ and $z$ directions. It is worth noting that this gives qualitatively different results than if we use a 2D square lattice model with periodic boundary conditions only in the $y$ direction. Since the 2D model does not have periodic boundary conditions in the $z$ direction, we do not get the $\sin(k_z )$ terms in Eq. \ref{Hamiltonian2} when considering a 2D square lattice. This makes the system invariant under $\phi$ rotations of $\boldsymbol{h}$. This implies that physical quantities such as $T_c$ and $F$ have the same angular dependence in the $xz$ and $yz$ plane, so that the system is not invariant under $\pi/2$ rotations in the $yz$ plane as is expected for a 3D cubic lattice. It should therefore be cautioned against simplifying the numerical simulations of a 3D cubic lattice by using a 2D square lattice model.
In our calculations we use $N_y = N_z$ so that we get an equal number of $k_y$ and $k_z$ values, thereby obtaining a $\pi/2$-rotational invariance in the $yz$-plane even when $N_y$ and $N_z$ are not much larger than the film thicknesses. It should also be noted that the thickness of the sample parallel to the interfaces is important for the physical results obtained in an experiment. In our paper, we have modeled a thin-film structure in which the width of the sample in the $y$ and $z$ directions is much larger than the thickness of the sample.

Before presenting our results, we finally also comment on the relevance of the BdG-lattice framework used here with respect to making predictions for experimentally realistic systems. The lattice framework has several advantages, such as capturing the crystal symmetry and its influence on physical quantities in addition to the fact that that the energy scales in the system can be varied across a large range. The main weakness with the present theoretical framework is that only relatively small sample sizes are computationally manageable, especially with periodic boundary conditions in two directions used here. When considering a thin superconducting layer, the superconducting coherence length must be short in order to be comparable to the thickness of the superconducting layer. $\xi$ is proportional to the inverse of the zero temperature gap of the superconducting layer. Considering a thin superconducting layer therefore results in a large value for the superconducting order parameter, and also a large critical temperature. However, the present framework can still be used to make qualitative and quantitative predictions for experimentally realistic systems, so long as the spatial dimensions are scaled by the superconducting coherence length. An example that illustrates that this method gives good agreement with experimental results when scaled in this way is Ref. \cite{Black-Schaffer2010Strongly}. This paper utilized the same theoretical formalism as we do here and the predictions made therein were later found to correspond very well to experimental measurements done in Ref. \cite{English2013Observation}. Thus, there is good reason to expect that the results obtained in the present framework for system parameters corresponding to a certain ratio between the system size $N_x/\xi$ should correspond well to experimental measurements on a system that has the same ratio between its length and the superconducting coherence length. This is the approach we will take below. 

\begin{figure}[t]
    \centering
    \includegraphics[width=\columnwidth]{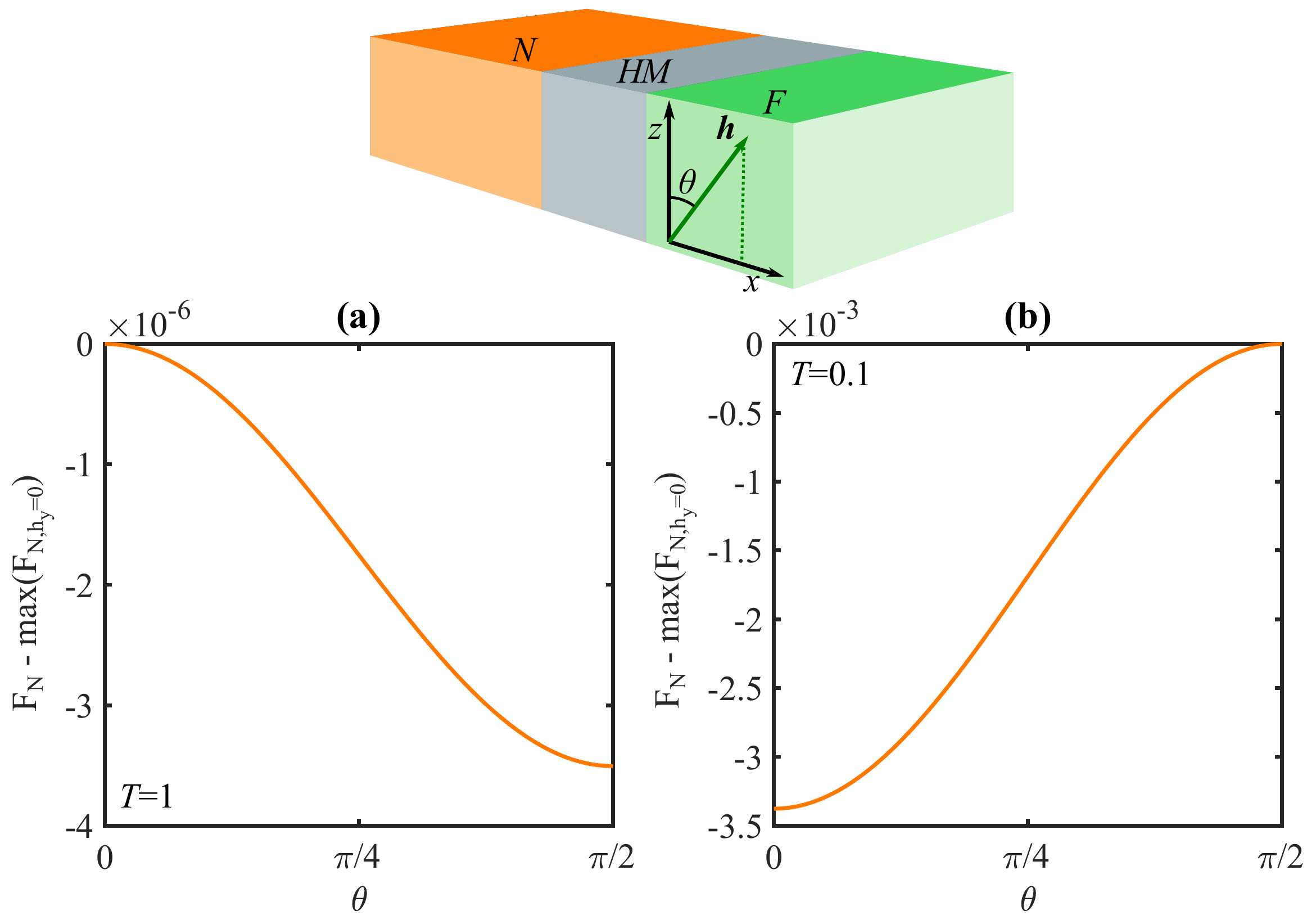}
    \caption{Panels (a) and (b) show $F_N (\theta)-\max(F_{N,h_y = 0})$ in the $xz$ plane for $T=1$ and $T=0.1$. The parameters used are specified in the main text. In (a) OOP magnetization is favored. In (b) IP magnetization is favored.}
    \label{fig:Fxz_diffT}
\end{figure}

\section{Results and Discussion}

\subsection{The non-superconducting contribution to the free energy}

\begin{figure}[t]
    \centering
    \includegraphics[width=\columnwidth]{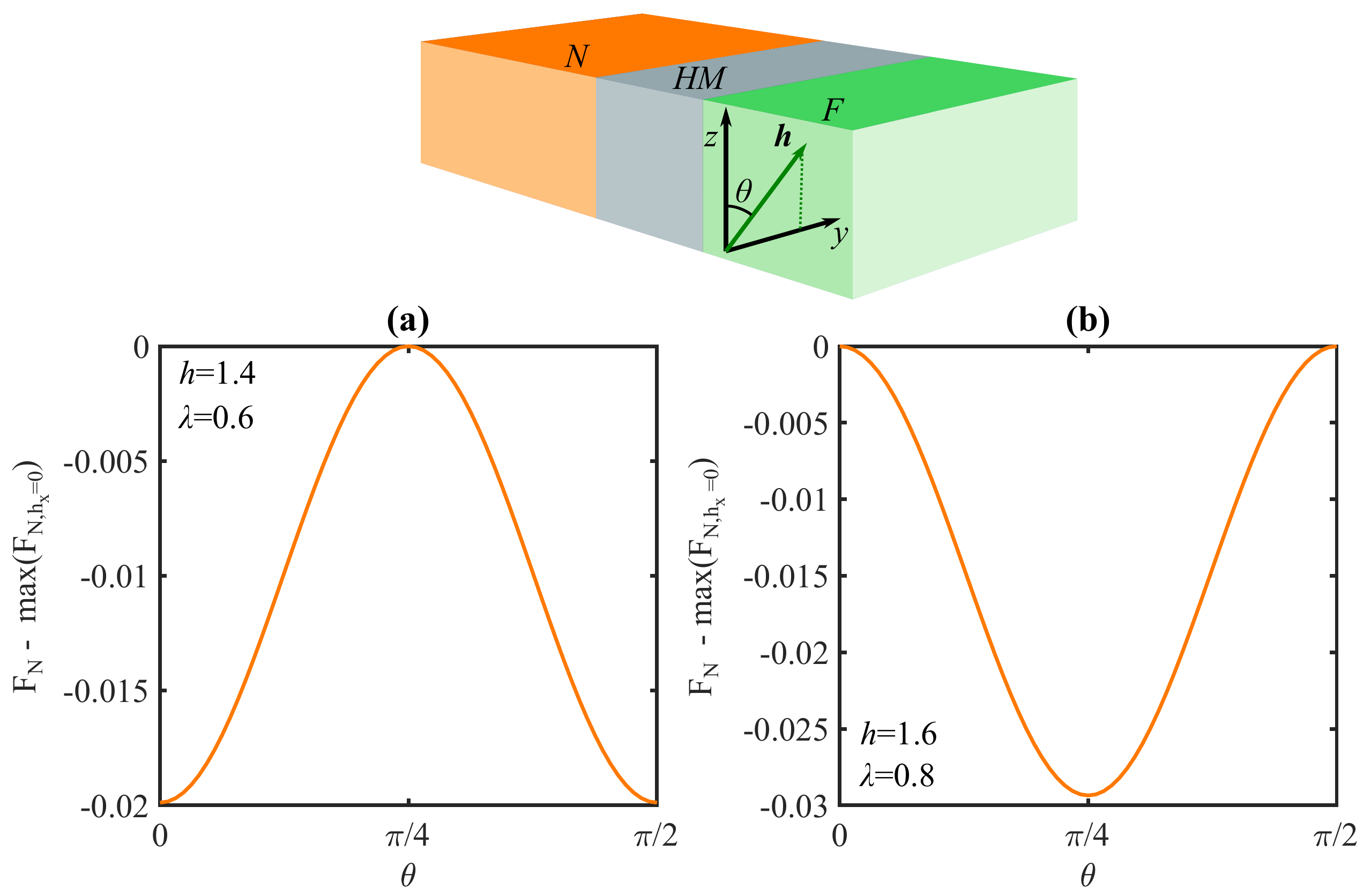}
    \caption{Panels (a) and (b) show $F_N (\theta)-\max(F_{N,h_x =0})$ in the $yz$ plane for $h=1.4$, $\lambda=0.6$ and for $h=1.6$, $\lambda=0.8$. The other parameters used are given in the main text. $T=0.01$. We see a $\pi/4$ rotation of the minimum from (a) to (b).}
    \label{fig:yzLambda}
\end{figure}

 We first look at a system as shown in Fig. \ref{fig:model}, where we have a normal metal $(N)$ rather than a superconductor, \textit{i.e.}, $U=0$. This is important in order to later distinguish the influence of the superconducting phase on the preferred magnetization orientation compared to the normal-state phase. We diagonalize the Hamiltonian described in Eqs. \ref{Hamiltonian} and \ref{Hamiltonian2} numerically using the parameters $N_{x,N}=9$, $N_{x,HM}=N_{x,F}=3$, $N_y = N_z =50$, $\mu_N =1.8$, $\mu_{HM} =1.7$, $\mu_F =1.6$, $h=1.4$ and $\lambda = 0.6$. We then plot the free energy for the N/HM/F trilayer, $F_{N}(\theta)$, to find the preferred direction of $\textbf{h}$ for a given $T$. In all free-energy plots we subtract the maximal free energy within the plane of rotation we are considering, \textit{i.e.} $\max(F_{N,h_y =0})$ when considering the $xz$ plane and $\max(F_{N,h_x =0})$ when considering the $yz$ plane. We do this to make it easier to compare the change in free energy for different parameter choices. Figure \ref{fig:Fxz_diffT} shows $F_N (\theta)$ in the $xz$ plane for $T=1$ and $T=0.1$. We see that the preferred magnetization direction may change as the temperature is increased. The preferred direction may also change when changing $h$, $\lambda$, or the layer thicknesses. The angular dependence of $F$ is the same for the $xy$ and $xz$ planes. Figure \ref{fig:yzLambda} shows $F_N(\theta)$ in the $yz$ plane at $T=0.01$ for different choices of $h$ and $\lambda$. We see that the preferred direction of $\boldsymbol{h}$ is rotated by $\pi/4$ when changing the parameters from $h=1.4$, $\lambda=0.6$ to $h=1.6$, $\lambda=0.8$. A similar rotation may also happen when changing $T$ or the layer thicknesses. Note that the free energy is invariant under a $\pi/2$ rotation in the $yz$ plane. This is reasonable, because a $\pi/2$ rotation of the cubic system around the interface normal should leave the system invariant independently of the magnetization direction. For sufficiently high temperatures, $F_N$ becomes constant. We underline that the effective magnetization anisotropy that arises here is distinct from the anisotropy terms described in Eqs. \ref{anis} and \ref{Keff}, the latter not being included in the analysis yet. We will shortly come back to the physical origin of the magnetic anisotropy in the present case. It is evident that the preferred direction of $\textbf{h}$ is highly dependent on the choice of parameters. To make a superconducting switch, we must therefore make sure that the non-superconducting contribution to the free energy favors a different magnetization direction than the superconducting contribution so that the superconducting and non superconducting contributions compete. We must also check whether a change in the preferred magnetization direction is actually caused by the superconducting contribution to $F$ and not by the non-superconducting contribution.
  
\begin{figure*}[t!]
    \centering
    \includegraphics[width=\textwidth]{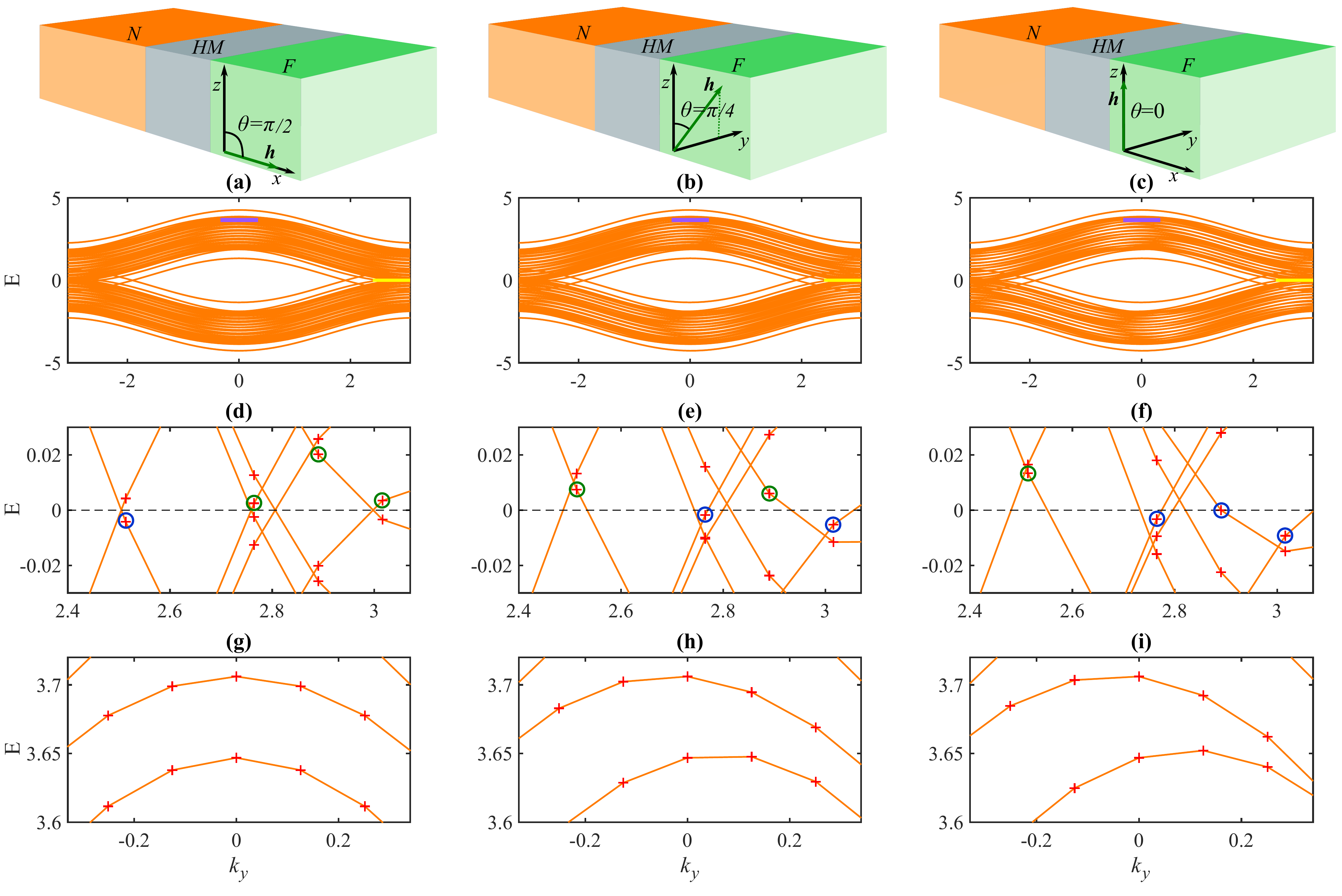}
    \caption{Panels (a)-(i) show the energy band structure, $E_{n,k_y ,k_z =0}(k_y )$, of the N/HM/S system. The parameters used are given in the main text. In (a), (d), and (g) we have OOP magnetization corresponding to the maximum of $F_N$ for low $T$. In (b), (e) and (h) we have IP magnetization with $\theta=\pi/4$ corresponding to the maximum IP value of $F_N$ for low $T$. In (c), (f), and (i) we have IP magnetization with $\theta=0$ corresponding to the minimum of $F_N$ for low $T$. Panels (d)-(f) show the band structure in the region marked in yellow in (a)-(c). The crosses mark the discrete eigenenergies. The encircled eigenenergies are shifted from above zero energy (green circle) to below zero energy (blue circle) or vice versa when rotating $\boldsymbol{h}$. At $T=0$ only eigenenergies below zero energy contribute to $F_N$. Panels (g)-(i) show the band structure in the region marked in purple in (a)-(c). Also for these higher-energy bands that contribute to $F$ at finite temperatures, there is a shift in the energy bands when rotating $\boldsymbol{h}$.}
    \label{fig:bandstructure}
\end{figure*}

\begin{figure}[t!]
    \centering
    \includegraphics[width=\columnwidth]{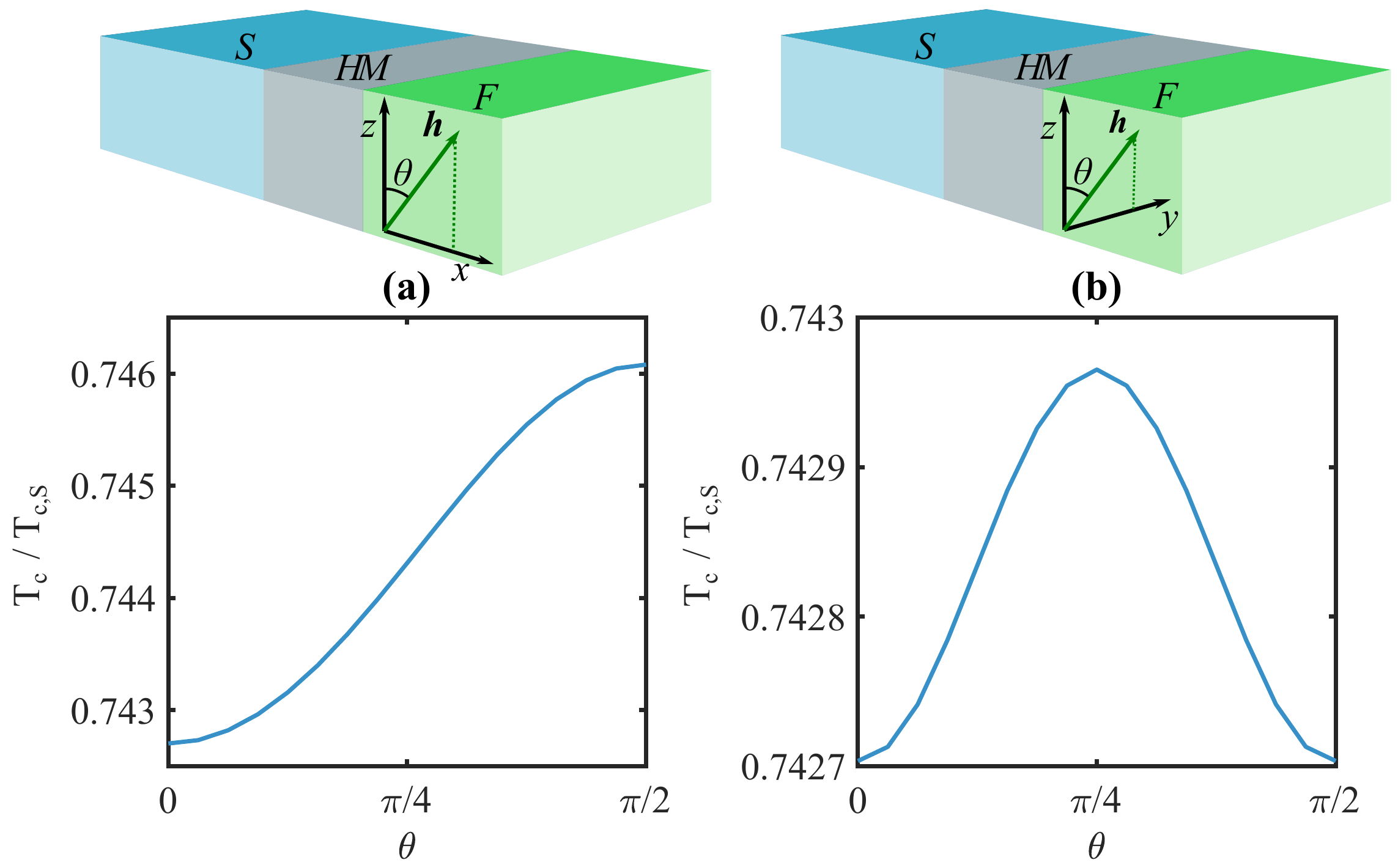}
    \caption{Panel (a) shows $T_c / T_{c,S} $ when rotating the magnetization from IP to OOP. Panel (b) shows $T_c / T_{c,S} $ for an IP rotation. The parameters used are specified in the main text.}
    \label{fig:Tc}
\end{figure}

Before turning to the superconducting case, we examine the energy band structure of the system in order to explain the change in free energy of the N/HM/F trilayer. If we consider small temperatures so that $F_N$ can be approximated by Eq. \ref{FT->0}, the free energy is determined by the sum over negative eigenenergies. If eigenenergies are shifted from above to below zero when some parameter is changed or if the eigenenergies below zero shift closer or farther away from zero, $F_N(\theta)$ will change. When increasing the temperature from zero, the smallest of the positive eigenenergies will give a contribution to the free energy. The band structure close to zero energy (relative the chemical potential) should therefore be of great importance to the free energy at low temperatures. In Fig. \ref{fig:bandstructure} we have plotted the energy bands, $E_{n,k_y ,k_z =0}(k_y )$, for three different magnetization directions. We consider the out-of-plane case $(\theta,\phi) = (\pi/2,0)$ and two in-plane cases $(\theta,\phi) = (\pi/4,\pi/2)$ and $(0,0)$, respectively. We have used the same parameters as in Fig. \ref{fig:Fxz_diffT}. In Fig. \ref{fig:bandstructure}, panels (a)-(c) show the overall band structure of the three magnetization directions. Panels (d)-(f) correspond to the region marked in yellow in (a)-(c) and show some of the eigenenergies close to zero energy. We see a variation in band structure between the different directions of $\boldsymbol{h}$. As a result some eigenenergies are shifted from above to below zero energy and vice versa. For $T\to0$ it is therefore likely that the differences in band structure cause the variation in $F_N$ for different magnetization directions. Note that this effective anisotropy is not caused by the discreteness of $k_y$ and $k_z$. In the limit where we have continuous energy bands, $N_y , N_z \to\infty$, the shifting of the energy bands should cause the same effect since finite sections of the continuous energy bands are shifted from above to below zero energy and vice versa. Panels (g)-(i) correspond to the purple region in (a)-(c) and show higher energy bands that only contribute to the free energy at finite temperatures. We see that the band structure has an angular dependence also at finite temperatures. It is therefore reasonable that $F_N$ has a temperature dependent angular dependence also for low, finite temperatures. For temperatures that are sufficiently high to make all energy eigenvalues partially occupied, $F_N(\theta)$ becomes gradually more independent of the magnetization direction.  Since $F_N(\theta)$ becomes constant for high temperatures, this indicates that the relative shift between the energy bands is such that it leaves the sum over all eigenenergies constant.

\begin{figure*}[t!]
    \centering
    \includegraphics[width=0.81\textwidth]{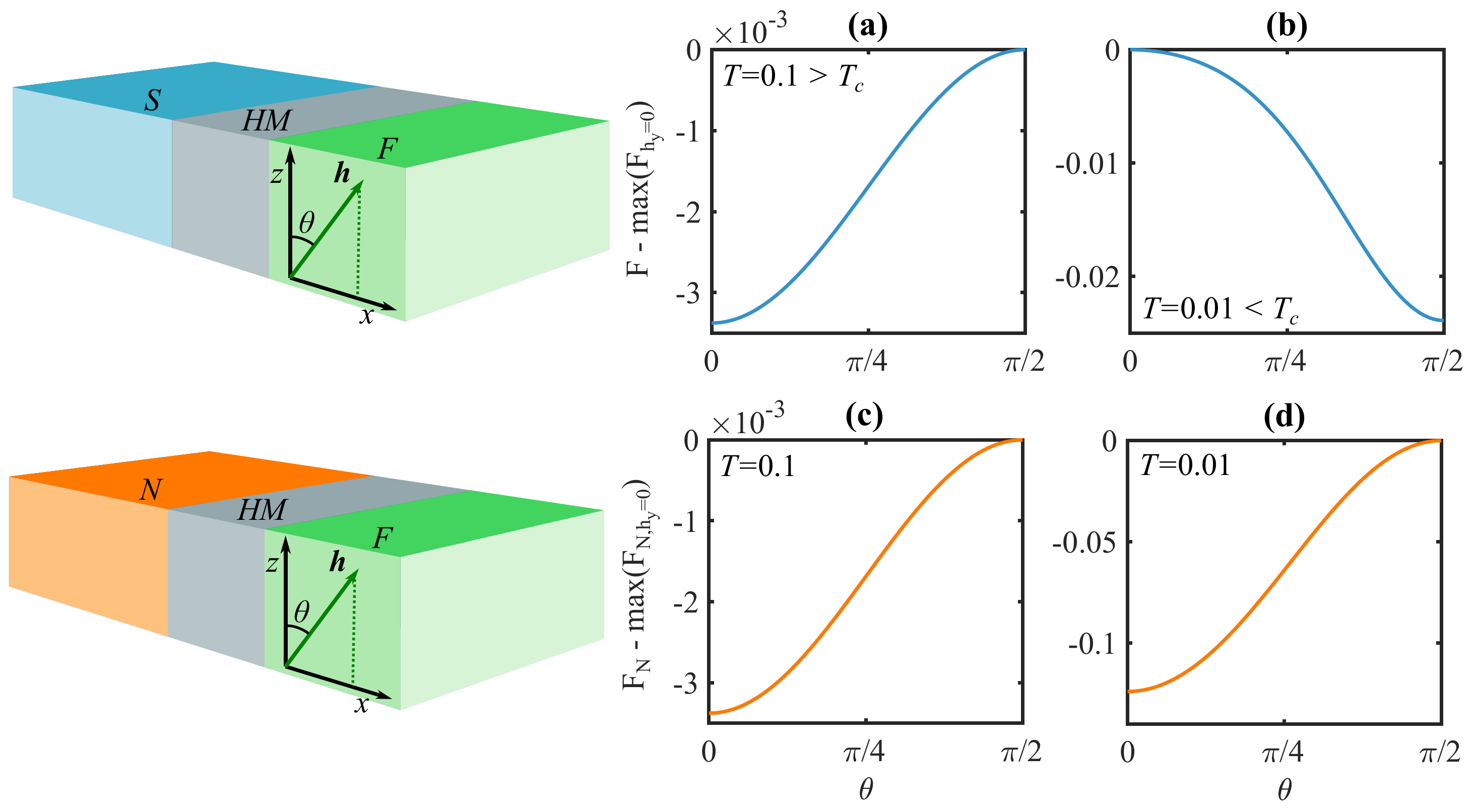}
    \caption{Panels (a) and (b) show $F (\theta)-\max(F_{h_y =0})$ in the $xz$ plane for $T=0.1>T_c=0.017$ and $T=0.01<T_c$, respectively. Panels (c) and (d) show the normal-state contribution to the free energy at the same temperatures. The parameters used are specified in the main text. From (a) and (b) we see that by decreasing the temperature below $T_c$ the preferred magnetization direction of the ferromagnet changes from IP to OOP. Since the normal-state contribution shown in (c) and (d) favors IP magnetization, the change in the preferred magnetization direction must be due to superconductivity.}
    \label{fig:SCswitch}
\end{figure*}
\begin{figure*}[t!]
     \centering
     \includegraphics[width=\textwidth]{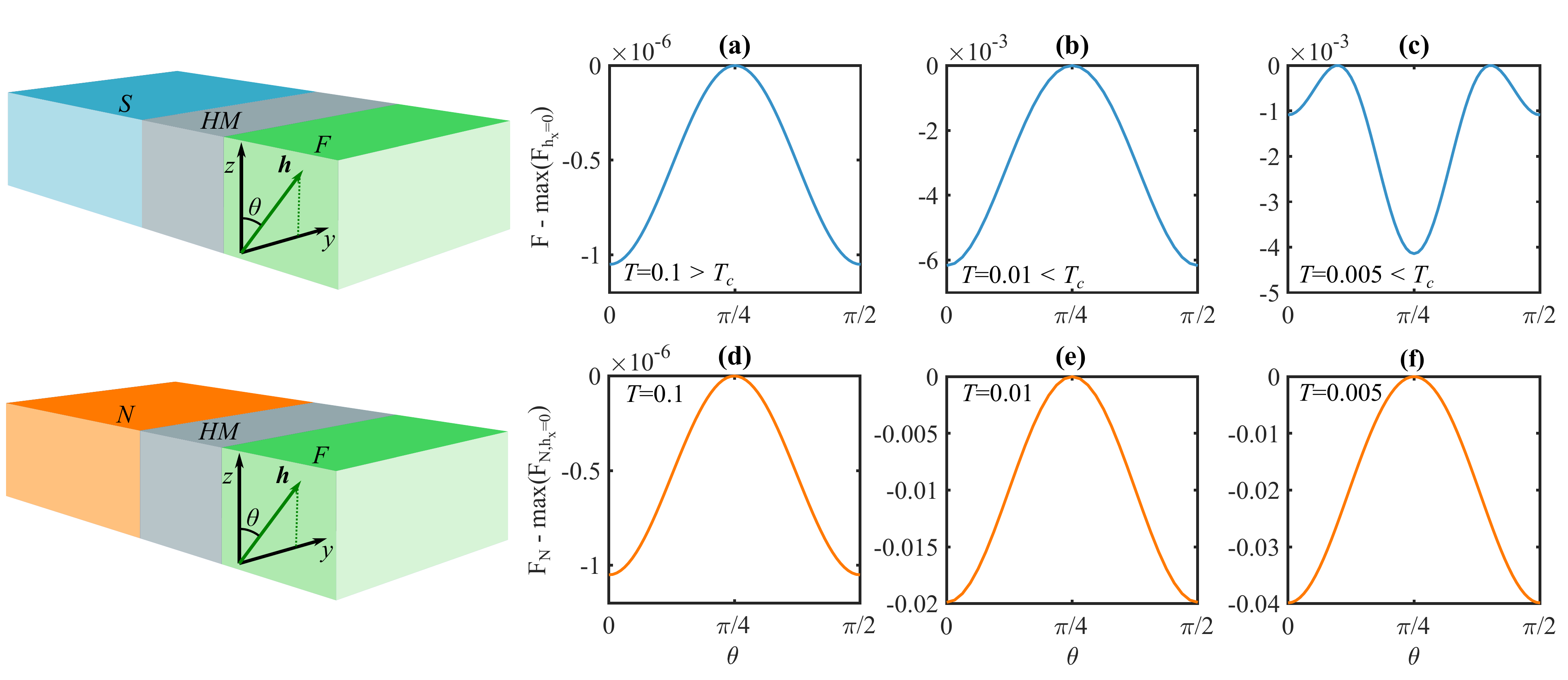}
     \caption{Panels (a), (b), and (c) show $F (\theta)-\max(F_{h_x =0})$ in the $yz$ plane for $T=0.1>T_c=0.017$, $T=0.01<T_c$ and $T=0.005<T_c$,  respectively. Panels (d), (e), and (f) show the normal-state contribution to the free energy at the same temperatures. The parameters used are specified in the main text. From (a), (b), and (c) we see that the IP minimum of the free energy rotates by $\pi/4$ at some temperature between $0.01$ and $0.005$, which are both below $T_c$. Since the normal-state contribution shown in (d), (e), and (f) favors magnetization along the crystal axes at all of these three temperatures, the change in the preferred magnetization direction must be due to superconductivity.}
     \label{fig:SCswitchyz}
\end{figure*}

\subsection{The superconducting contribution to the free energy}
We now look at a system as shown in Fig. \ref{fig:model}, where we have a superconductor, \textit{i.e.}, $U>0$. The basic question we seek to address is, is it possible to trigger a reorientation of the preferred magnetization direction in the system via a superconducting phase transition, \textit{i.e.}, by adjusting the temperature from above to below $T_c$? We diagonalize the Hamiltonian described in Eqs. \ref{Hamiltonian} and \ref{Hamiltonian2} numerically using the parameters $N_{x,S}=9$, $N_{x,HM}=N_{x,F}=3$, $N_y = N_z =50$, $\mu_S =1.8$, $\mu_{HM} =1.7$, $\mu_F =1.6$, $U=1.9$, $h=1.4$ and $\lambda = 0.6$. For this parameter set the superconducting coherence length is $\xi=5$. We expect that our results can be generalized to systems with thicker layers as long as the relative thicknesses of the layers compared to the coherence length stay constant, as explained previously in this paper.
\begin{figure*}[t!]
    \centering
    \includegraphics[width=\textwidth]{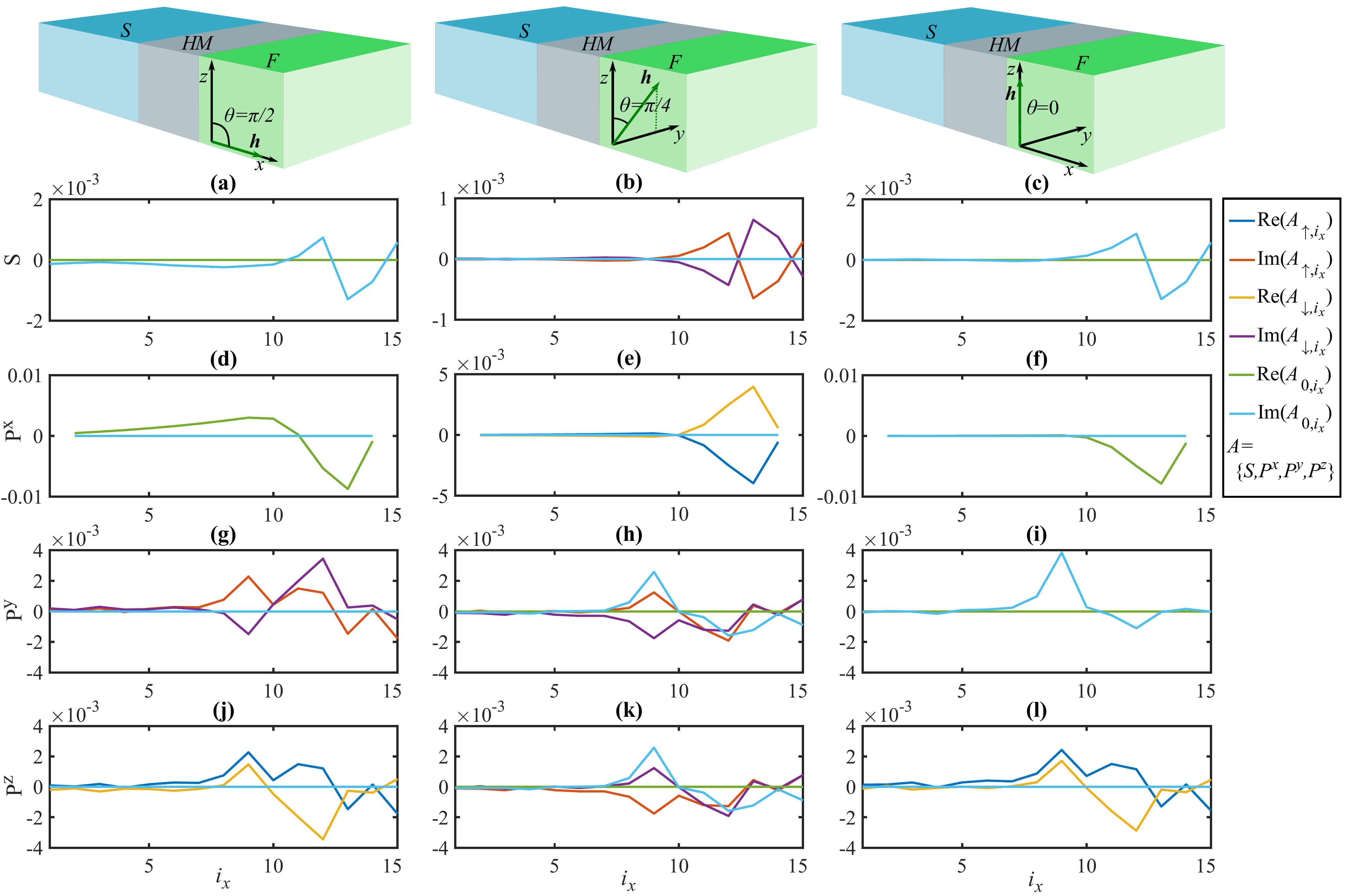}
    \caption{Panels (a)-(l) show the triplet amplitudes generated in the S/HM/F system at $T=0.01 < T_c =0.017$. The parameters are given in the main text. Panels (a)-(c) show the $s$-wave triplet amplitudes, (d)-(f) show the $p_x$-wave triplet amplitudes, (g)-(i) show the $p_y$-wave triplet amplitudes and (j)-(l) show the $p_z$-wave triplet amplitudes. The first column corresponds to OOP magnetization, the second column corresponds to IP magnetization with $\theta=\pi/4$, and the third column corresponds to IP magnetization with $\theta=0$. Note that all amplitudes that are not visible in the plots are either zero or close to zero.}
    \label{fig:triplets}
\end{figure*}

We begin by considering the dependence of the superconducting critical temperature on the magnetization direction. Since we have chosen a size of the superconductor larger than the coherence length, the magnitude of the change in critical temperature is rather small. $T_c(\theta)/T_{c,S}$ is plotted in Fig. \ref{fig:Tc}. $T_{c,S}$ is the critical temperature of the superconducting layer without the heavy-metal layer and the ferromagnetic layer. In (a) we see a suppression of $T_c$ for IP magnetization as found by experiments \cite{Banerjee2017Controlling} on a similar system. Panel (b) shows an additional IP variation in $T_c$, where $T_c$ is suppressed along the cubic axes. In our system, where the thickness of the superconductor is less than twice the coherence length, we do not obtain a substantial bulk region with a constant gap. When calculating $T_c$ we measure the change in the gap in the middle of the superconducting region when recalculating the gap $m$ times. This means that for superconducting layers that are not much longer than the coherence length, our method for calculating $T_c$ is not entirely accurate unless $m$ is chosen to be very large. Therefore, we set $m=150$. The change in $T_c$ when increasing $m$ by 10 is then $10^{-4}T_{c,S}$, which is a small change compared to the total change in $T_c$ when rotating $\boldsymbol{h}$. We have checked that we get a qualitatively similar behavior of $T_c$ to that in Fig. \ref{fig:Tc} for thicker superconducting layers. $T_{c,S}$ was calculated with $m=200$. The number of times we divided our temperature interval is $n=20$, making $m$ the parameter that restricts the accuracy of our $T_c$ calculation. The reason we chose a superconductor of only 9 lattice points is that a long superconducting layer requires a low $U$ to obtain a coherence length that is comparable to the thickness of the superconducting layer. This results in a very low critical temperature. At very low temperatures only the eigenenergies below zero contribute to the free energy as shown in Eq. \ref{FT->0}. If we have few $k_y$ and $k_z $ values, the shifting of eigenenergies from above to below zero energy will have a great impact on the free energy. This is especially a problem when computing the non-superconducting contribution to the free energy, where we have no gap and many eigenenergies are close to $E=0$. We therefore do not get a smooth curve when plotting $F_N (\theta)$. To avoid this problem we must either choose a short superconductor such that we can look at higher temperatures, or let $N_y$ and $N_z$ be very large. The latter option makes the free-energy calculations computationally expensive, which is why we chose the former. Note that we would expect a stronger variation in $T_c$ if we made our superconductor comparable to the coherence length rather than almost two times larger. 

From the angular dependence of $T_c$ we may expect a superconducting contribution to the free energy in which $F$ is increased for the IP orientation, especially along the cubic axes. Figure \ref{fig:SCswitch} shows the free energy in the $xz$ plane for $T=0.1>T_c$, $T=0.01<T_c$ and $T=0.005<T_c$. As expected, we see a change in the preferred magnetization direction due to the fact that the superconducting contribution to $F$ favors OOP magnetization while the non-superconducting contribution to $F$ favors IP magnetization. Figure \ref{fig:SCswitchyz} shows the free energy in the $yz$ plane for the same temperatures. For sufficiently low $T$, the superconducting contribution to the free energy starts to dominate, and we have an IP $\pi/4$ rotation of the minimum of free energy. Notice however that the IP variation in the free energy is weaker than the IP-OOP variation. Therefore OOP magnetization is favored as the ground state of the system despite the fact that the free energy also varies when the magnetization is rotated IP. For both the $xz$ and $yz$ planes the change in preferred magnetization direction will generally occur at lower temperatures than $T_c$, meaning that the superconducting contribution does not necessarily start to dominate exactly at the critical temperature. When increasing $T$ the preferred magnetization direction at some point changes from IP to OOP without any involvement of superconductivity. This is exemplified by the behavior of $F_N(\theta)$ in Fig. \ref{fig:Fxz_diffT}, which was plotted for a temperature $T > T_c$. The superconducting switch must therefore be operated over a limited temperature range around the temperature at which the change in the preferred magnetization direction occurs. However, we discuss toward the end of this paper how the superconducting contribution to the free energy, causing an effective magnetic anistropy, can be experimentally detected even in the cases in which the superconducting contribution is not sufficiently strong to change the preferred magnetization orientation.

\begin{figure}[t!]
    \centering
    \includegraphics[width=\columnwidth]{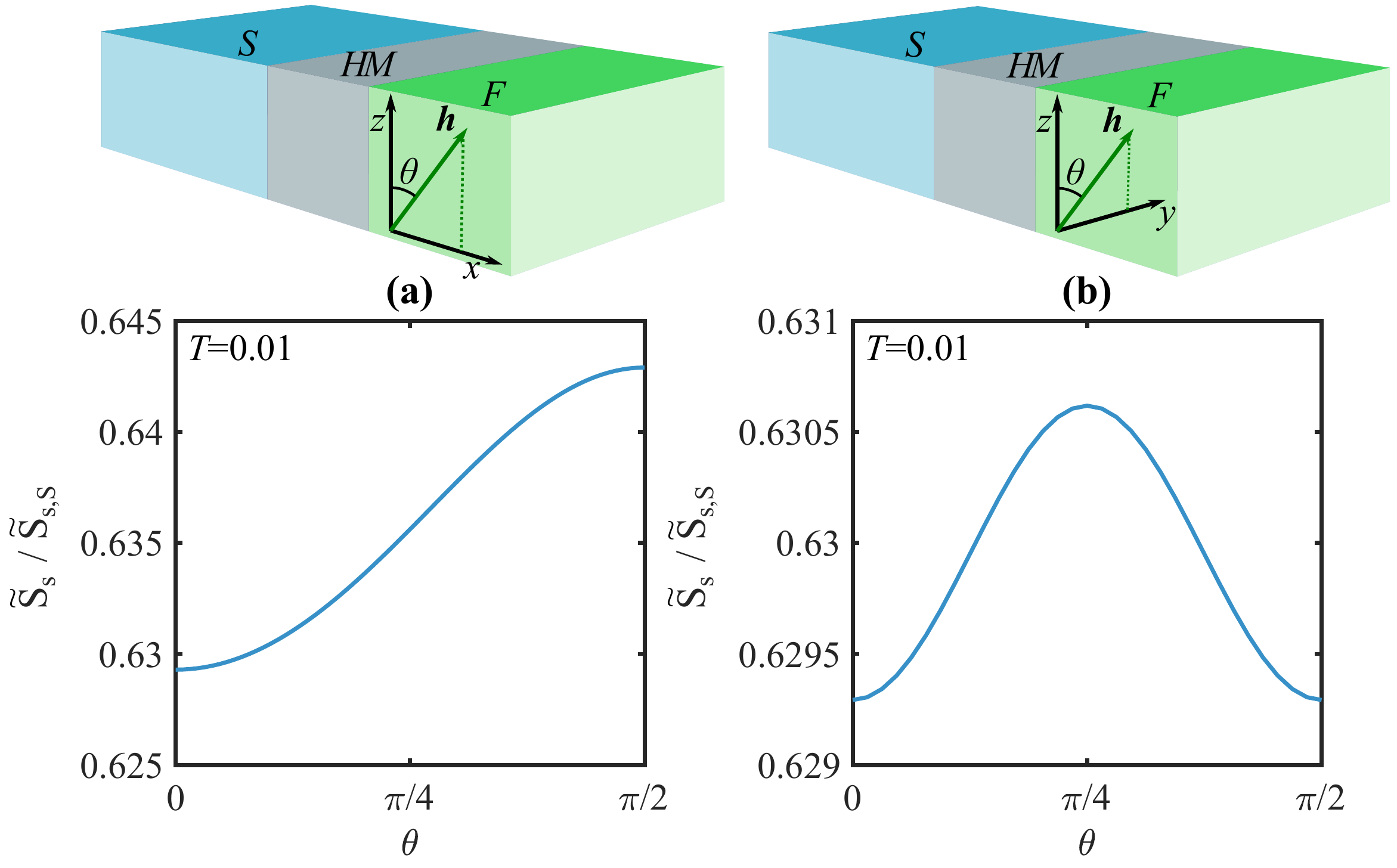}
    \caption{Panels (a) and (b)  the singlet amplitude $\Tilde{S}_s/\Tilde{S}_{s,S}$ in the $xz$ and $yz$ plane. The parameters are given in the main text. $T=0.01 < T_c =0.017$.}
    \label{fig:singlets}
\end{figure}

The angular dependence of $T_c$ and of the superconducting contribution to $F$ can be explained by the generation of triplet Cooper pairs. At an S/F interface, the spin splitting of the energy bands of the ferromagnet causes transformation of singlet Cooper pairs into opposite-spin triplets. The Rashba spin-orbit coupling terms in the Hamiltonian in Eqs. \ref{Hamiltonian} and \ref{Hamiltonian2} are proportional to $\sin(k_y)$ and $\sin(k_z)$. Therefore, electrons experience different energies if the sign of $(k_y ,k_z )$ is changed. This symmetry-breaking causes triplet generation at the S/HM interface, and enables equal-spin triplet generation, depending on the relative orientation of the magnetization and the spin-orbit field. In Fig. \ref{fig:triplets} we have plotted the triplet amplitudes corresponding to OOP magnetization and the IP magnetization directions $(\theta,\phi)=(\pi/2,\pi/4)$ and $(0,0)$, respectively. The relative time used in the computation of the s-wave odd-frequency triplet amplitudes is $\tau=5$. We see that there is a generation of short-range and long-range triplet amplitudes depending on the magnetization direction. The generation of triplet amplitudes lowers the singlet amplitude in the superconductor, since singlet Cooper pairs are converted into triplet Cooper pairs. In Fig. \ref{fig:singlets} we have plotted $\Tilde{S}_s/\Tilde{S}_{s,S}$, where $\Tilde{S}_s$ is defined in Eq. \ref{S}. $\Tilde{S}_{s,S}$ is the singlet amplitude in the superconducting layer without the heavy-metal layer and the ferromagnetic layer. We see that the singlet amplitude  is suppressed for IP magnetization, especially along the cubic axes.  Since the singlet amplitude is proportional to the superconducting order parameter, a suppression of the singlet amplitude should lead to a decrease in $T_c$ and an increase in $F$. This is exactly what we have seen from Figs. \ref{fig:Tc}, \ref{fig:SCswitch} and \ref{fig:SCswitchyz}. We may therefore explain the variation in $T_c$ and $F$ by the generation of triplet amplitudes depending on the relative orientations of the spin-orbit field and the magnetization. 

The diffusive limit calculations in Ref. \cite{Banerjee2017Controlling} found an IP suppression of $T_c$ as in our calculations. However, in the diffusive limit $T_c$ was found to be invariant under IP rotations of the magnetization. In Ref. \cite{Banerjee2017Controlling} the HM/F layer is modelled as a single layer with the exchange field and the spin-orbit coupling as homogeneous background fields, which similarly to what occurs in the ballistic limit results in a generation of both short-range and long-range triplets close to the interface. The IP suppression of $T_c$ compared to $T_c$ at OOP magnetization is both for the ballistic and the diffusive limit a result of differences in the triplet generation when the exchange field is parallel and perpendicular to the interface between the superconductor and the HM/F layer. The change in $T_c$ under IP rotations of the magnetization found in the present paper is a result of differences in the triplet generation at different IP magnetization directions due to the crystal structure of the lattice in the HM region. This is the reason why these variations are not found in the diffusive limit calculations in Ref. \cite{Banerjee2017Controlling}, which does not model the S/HM/F system by a lattice model. For very thin films, like the ones considered experimentally in Ref. \cite{Banerjee2017Controlling}, we expect the sample to approach the ballistic limit such that a variation in $T_c$ for IP rotations of the magnetization should be observable.

\subsection{The shape anisotropy contribution to the free energy}

\begin{figure}[t!]
    \centering
    \includegraphics[width=0.9\columnwidth]{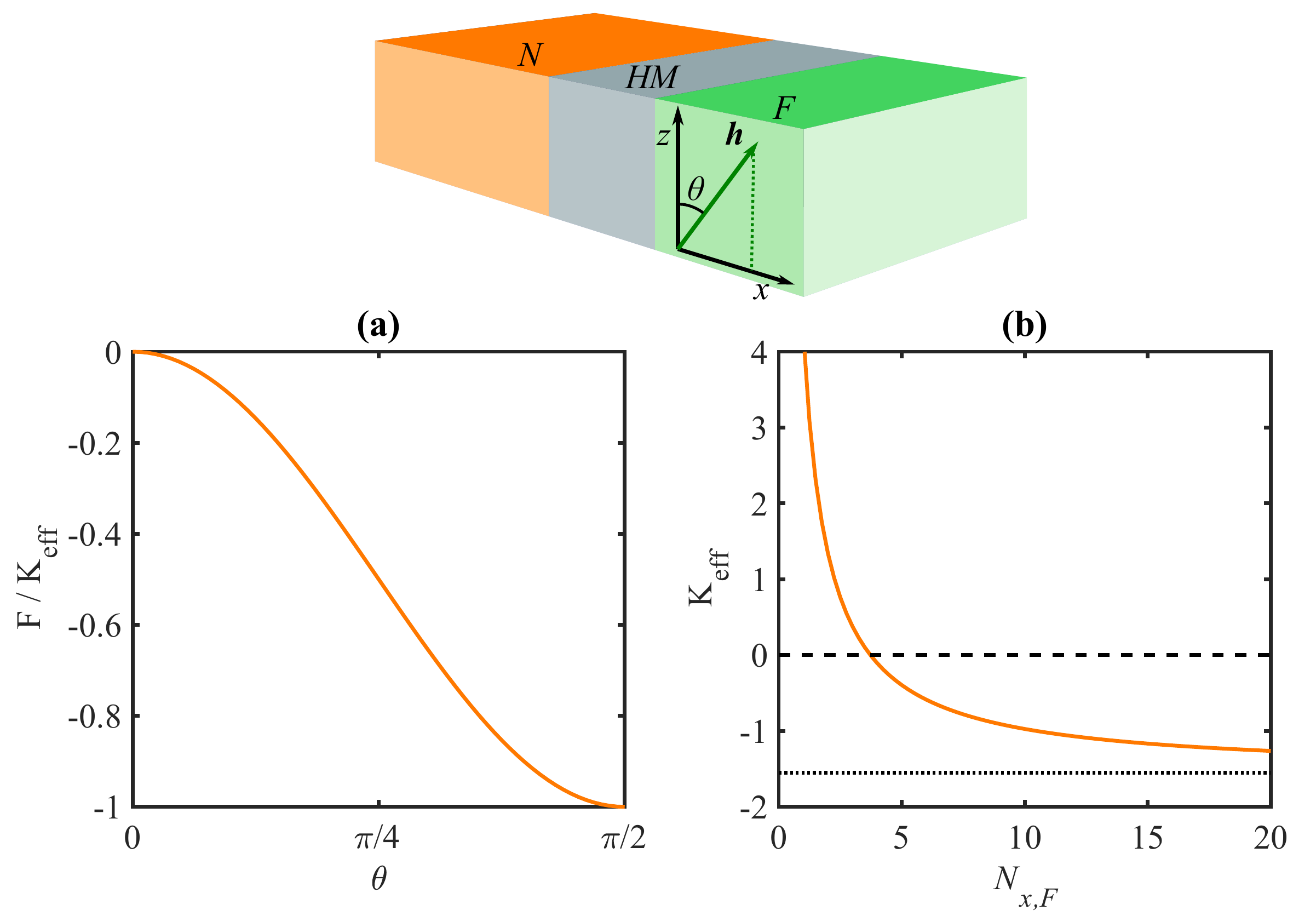}
    \caption{Panel (a) shows the perpendicular/shape anisotropy contribution to $F$. Panel (b) shows the effective anisotropy constant, $K_{eff}$, as a function of the ferromagnet thickness.}
    \label{fig:Anisotropy}
\end{figure}

Until now, we have disregarded the intrinsic magnetic anisotropy of the thin ferromagnetic film, which does not depend on the coupling to itinerant electrons $\{c,c^\dag\}$ in our model. For concreteness, we will now consider the case of a Pt heavy-metal layer and a Co(111) ferromagnetic layer. In this case, the anisotropy constants are \cite{Johnson1996Magnetic} $K_v =-0.77$ MJ/m$^3$, $K_i = 1.15$ mJ/m$^2$ and $K_s = -0.28$ mJ/m$^2$. The lattice constants of Co are \cite{Ullmann2003} $a_x =a_y =251$pm and $a_z =407$pm. The anisotropy contribution to the free energy is plotted in Fig. \ref{fig:Anisotropy}a for this choice of parameters. The effective anisotropy constant defined in Eq. \ref{Keff} is plotted in Fig. \ref{fig:Anisotropy}b as a function of $N_{x,F}$. By solving $K_{\text{eff}}=0$, we find that the anisotropy contribution to the free energy favors an OOP magnetization for $N_{x,F} \leq 3$ and an IP magnetization for $N_{x,F} \geq 4$. Since we may generalize our results to any system size as long as the layer thickness relative to $\xi$ stays constant, we may consider a system with any $N_{x,F}$. By making the ferromagnetic layer thick enough to give a contribution to $F(\theta)$ favoring an IP magnetization, but thin enough that $K_{\text{eff}}$ is small, it is in principle possible to get an IP-OOP superconducting switch despite the fact that the non superconducting contribution to $F$ has gained an extra term. We may also make the ferromagnetic layer so thick that the non-superconducting contribution to the free energy enforces IP magnetization. Since the shape anisotropy contribution to the free energy is invariant under rotations in the $yz$ plane, we may get a $\pi/4$ rotation in the magnetization as shown in Fig. \ref{fig:SCswitchyz}. This means that an IP superconducting switch in the magnetization direction is in principle possible, even if the preferred magnetization direction is OOP when disregarding shape anisotropy. The possibility of changing the preferred direction in the $yz$ plane is interesting as the magnetic field of the ferromagnet in such a case is not perpendicular to the superconducting layer. We therefore avoid demagnetising currents close to the interface in the superconducting region as well as vortex formation inside the superconductor \cite{Fossheim2004Superconductivity}. For magnetization with an OOP component, demagnetization effects may be of greater importance. 

It is worth noting that even if the ferromagnetic layer is so thick that the non-superconducting contribution dominates, it may still be possible to measure the superconducting contribution to the free energy. The superconducting contribution to the free energy in an F1/S/F2 system can be measured \cite{Zhu2016Superconducting} by applying an external magnetic field and measuring the critical field needed to flip the magnetization from an antiparallel to a parallel alignment. It should be possible to do similar measurements on the S/HM/F-system. For instance, one could apply an external field to flip the magnetization of the ferromagnet between the IP and the OOP direction. The superconducting contribution favors OOP magnetization and would therefore reduce the critical field needed to flip the magnetization from IP to OOP orientation. Such a reduction of the critical field would thus be evidence of a superconductivity-induced anisotropy term for the ferromagnet.

\section{Concluding remarks}

This work predicts a possible reorientation of the magnetization direction of a thin-film ferromagnet upon lowering the temperature below the superconducting critical temperature $T_c$ when the ferromagnet is separated from a superconductor by a thin heavy-metal film. Especially for a thin ferromagnetic film with weak shape anisotropy, the superconducting phase transition should induce an in-plane to out-of-plane rotation of the magnetization. We have also found that if the shape anisotropy is strong enough to enforce an in-plane magnetization direction, a $\pi/4$ in-plane rotation of the magnetization can occur upon lowering the temperature below $T_c$. In addition, we have considered the dependence of $T_c$ on the magnetization direction. Here, we find that our lattice-model calculations predict an additional in-plane variation in $T_c$ compared to the previous diffusive-limit calculations, which only show an in-plane suppression of $T_c$ independently of the in-plane magnetization orientation. Both the $T_c$ suppression and the magnetization reorientation can be explained by the generation of short-range and long-range triplet Cooper pairs close to the interfaces depending on the relative orientations of the exchange field of the ferromagnet and the spin-orbit field of the heavy metal. Our results should be reproducible experimentally for systems with the same ratio between the layer thicknesses and the superconducting coherence length.

\acknowledgments

We thank J. A. Ouassou, V. Risingg{\aa}rd, M. Amundsen, and M. G. Blamire for helpful discussions. This work was supported by the Research Council of Norway through its
Centres of Excellence funding scheme grant 262633 “QuSpin”. N.B. was supported by the UKIERI grant from the British Council.

\end{document}